\documentclass[aps,prl,twocolumn,showpacs]{revtex4}
\usepackage{graphicx}
\usepackage{bm}
\input colordvi
\begin{document}
\title{Nonlinear current response of an isolated system of interacting fermions}
\author {Marcin Mierzejewski $^{1,2}$ and Peter Prelov\v{s}ek $^{1,3}$ }
\affiliation{ $^1$ J. Stefan Institute, SI-1000 Ljubljana, Slovenia }
\affiliation{$^2$ Institute of Physics, University of Silesia, 40-007 Katowice, Poland}
\affiliation{$^3$ Faculty of Mathematics and Physics, University of Ljubljana, SI-1000 Ljubljana, Slovenia }

\begin{abstract}
Nonlinear real--time response of interacting particles is studied on the example of 
a one--dimensional tight--binding model of spinless fermions driven by electric field.
Using equations of motion and numerical methods we show that for a non--integrable case
at finite temperatures the major effect of nonlinearity can be taken into account 
within the linear response  formalism extended by a renormalization of the kinetic energy 
due to the Joule heating.  On the other hand, integrable systems show on constant driving 
a different universality  with a damped oscillating current whereby the frequency is related 
but not equal to the Bloch oscillations. 
\end{abstract}


\maketitle
Despite its importance for various branches of physics, 
the real--time response of quantum systems remains 
in many aspects an unexplored field. Only recently have the 
time--resolved measurements provided important information
on the nonequilibrium short--time dynamics of correlated bulk materials \cite{bulk_e}
nanostructures \cite{nano_e} and optical lattice systems \cite{opt_e}. 
In contrast
to the developments in the experimental methods,  
theoretical description of the real--time dynamics remains a difficult 
and challenging task. As the exact time evolution is known only
for very few models (see the discussion in Ref. \cite{demler}), 
 most of unbiased results has been obtained from advanced 
numerical approaches like  exact diagonalization (ED) \cite{ED}, 
time--dependent density matrix renormalization group (tDMRG) \cite{DMRG} 
or nonequilibrium dynamical mean--field theory (nDMFT) \cite{DMFT}. 
These approaches allow for studying various phenomena, e.g., 
the nonlinear transport through  interacting nanosystems \cite{qd1} ,
the electric--field induced breakdown of the Mott insulator \cite{ED,mt1},
or Bloch oscillations \cite{DMFT,zoto}. In most cases, theoretical predictions for the real--time response
can be formulated only on the basis of numerical studies. 

 It is understood that for a weak driving force, the real--time response
is determined by the equilibrium correlation functions as described
by the linear--response (LR) theory. This theory has recently been extended 
to account for a system, that was previously driven out of equilibrium \cite{kollar}.
 Furtheron we focus on the case of 
the charge current $I(t)$ in a one-dimensional (1D) system as induced by 
electric field $F(t)$, 
whereby the relevant equilibrium LR function is the dynamical conductivity $\sigma(\omega)$.  
For a generic (non--integrable)  system of interacting  (correlated) 
fermions at finite temperature $T>0$ 
one expects  within the  LR regime the  relaxation of current 
due to (Umklapp) scattering processes and hence finite d.c. value $\sigma_0=\sigma(0)$.  
This, in turn, leads to a steady current  under a constant driving field $F$.
In contrast, it has been recognized that integrable systems in spite of interaction reveal
a dissipationless component of the current response even at $T>0$ as manifested 
by a finite charge stiffness $D(T>0)>0$ \cite{u2}. Hence, the qualitative difference between 
both types of systems is expected to remain even for the nonlinear transport.

In this Letter, we address two aspects of the nonlinear transport in 
an isolated  system of interacting tight--binding fermions under a time--dependent driving
force: ({\em i}) How to generalize the LR response approach of generic systems at $T>0$ 
to stronger  fields and longer steady driving? Here we show that
the dominating lowest--order 
mechanism beyond the LR regime is the increase of internal system energy 
or Joule--type heating 
(although we are not dealing with relaxation to canonical equilibrium)  
proportional to the square of the electric field. It could be accounted for by renormalization of 
the kinetic energy (as the sum rule for $\sigma(\omega)$ and, consequently the  
relevant  $\sigma_0$). This extension allows one to predict 
strongly nonlinear response {\em without explicit solution} of the von Neumann 
or the time--dependent Schr\"odinger equations.
({\em ii}) Is there a qualitative difference in nonlinear response between 
integrable and non--integrable systems? Our results reveal a clear distinction
between both categories  whereby the response of integrable system to a steady field 
$F(t>0)= \mathrm{const}$ resembles the (damped) Bloch
oscillations of noninteracting (NI) fermions, however, with nontrivially modified
oscillation frequency.

We investigate an isolated 1D system of charged spinless fermions 
with periodic boundary conditions. The system is threaded by  a 
time--dependent magnetic flux  
$\tilde \phi(t)$ and we assume that the flux enters only the kinetic energy term 
$H_k$ of the Hamiltonian $H$ 
\begin{eqnarray}
H&=& H_k+ H_I, \nonumber \\
H_k&=&-t_h \sum_j \left\{ {\mathrm e}^{i \phi(t)}\; c^{\dagger}_{j+1}c_j +{\mathrm h.c.} \right\},
\label{h1}
\end{eqnarray}
where  $\phi=\tilde \phi/L$ is the flux per bond, $L$ is the number of sites and $H_I$ 
is assumed to be flux independent. We put furtheron $t_h=1$. The time--dependent flux induces an electric field 
$F=- \dot{\phi}$.    The charge current operator can be written as 
\begin{equation}
J= -\frac{\partial H}{\partial \tilde \phi} =\frac{i}{L} \sum_j \left\{  {\mathrm e}^{i  \phi(t)}  \; 
c^{\dagger}_{j+1}c_j -{\mathrm h.c.} \right\}.
\label{jdef}
\end{equation} 

Before specifying a particular form of the interaction term $H_{I}$ and discussing numerical results 
it is instructive to derive simple equations of motion for the total energy 
$E(t)=\langle H(t) \rangle ={\mathrm Tr}[\rho(t) H(t)]$, the kinetic energy  
$E_k(t)= \langle H_k(t) \rangle$ and the current $I(t)=\langle J(t) \rangle$.
Making use of the von Neumann equation $ i  \dot{\rho}(t)= [H(t),\rho(t)] $ 
one can easily find the relations
\begin{eqnarray}
\dot{E}(t)&=& -i {\mathrm Tr}\left\{ \left[H(t),\rho(t)\right] H(t)\right\}+
\dot{\phi}(t) {\mathrm Tr} \left[ \rho(t) \frac{\partial H}{\partial \phi} \right] \nonumber \\
&=& L\;F(t)\;I(t), \label{hdot} \\
\dot{E}_k(t)&=& i \langle[H(t),H_k(t)]\rangle+  L\;F(t)\;I(t), \label{hkdot} \\
\dot{I}(t)&=& i \langle[H(t),J(t)]\rangle-  F(t) \frac{E_k(t)}{L}. \label{idot}
\end{eqnarray} 
In the case of NI fermions $H_I=0, H=H_k$ and commutators in 
Eqs. (\ref{hkdot}-\ref{idot}) vanish. Then, for a constant electric field $F(t)=F$  equations 
lead to harmonic oscillations with a frequency $\omega_{B}=  F $, the solution known 
as Bloch oscillations.

Eq.~(\ref{idot})  offers also a nontrivial approach to the nonlinear inverse problem, i.e., 
for arbitrary  interaction term $H_I$ one can find the tuning of the electric pulse $F(t)$ so 
that the induced current follows required time profile $I_{\rm a}(t)$.
Namely, putting $I_{\rm a}(t)$ in the lhs. of Eq.~ (\ref{idot}) one can solve the equation for $F(t)$. 
This value of $F(t)$ determines the magnetic flux that should be used in the subsequent time step of
the numerical solution of the von Neumann equation. 
Numerical results based on this method will be presented elsewhere. 

In the following we study numerically the  real--time current response within the  1D $t-V-W$ model,
\begin{equation}
H_I=V \sum_j \hat{n_j} \hat{n}_{j+1} +W \sum_j \hat{n_j} \hat{n}_{j+2} ,
\end{equation}
where $\hat{n}_j= c^{\dagger}_{j}c_j$, whereas $V$ and $W$ are (repulsive) interactions
among particles on the nearest--neighbor and next--nearest--neighbor sites, 
respectively. In the absence of the external field the $t-V$ ($W=0$) model is integrable
and within the LR theory $\sigma(\omega)$ has anomalous properties \cite{u2}. 
In particular, it can show a dissipationless component even at $T>0$ , 
i.e. $\sigma(\omega \sim 0) = D(T) \delta(\omega)$.  $W \neq 0$ breaks  the integrability 
leading to a finite d.c. $\sigma_0(T>0) < \infty$  \cite{u2}. In the following we 
focus on  the {\em metallic} regime of  the half--filled systems with $n=N_e/L
=1/2$ and $V < 2t_h$.
 
Our aim is to study primarily a generic situation at $T>0$ to avoid more specific cases 
emerging from the ground state. We perform the numerical evolution of the many-fermion 
wavefunction $|\Psi_l(t) \rangle$. 
The initial $|\Psi_l(0) \rangle$ should in principle be chosen as eigenstates of 
$H(0)$ distributed according to the canonical ensemble. Since the latter approach is 
possible only via full ED (typically $L<16$ for the problem
under consideration), we instead work using 
the microcanonical Lanczos method (MCLM) \cite{mclm} allowing systems up to $L=26$.
First, we numerically generate approximate $|\Psi_l(0) \rangle , l= 1,N_s$
with energy $E(0)= \langle H(0) \rangle$ corresponding to the canonical value for given $T$ (and $L$),
but as well with a small energy uncertainty $\delta^2 E=\langle [H(0)-E(0)]^2 \rangle$.  
The time evolution $|\Psi_l(t) \rangle$ is then calculated by step-vise change of $\phi(t)$ in small 
time increments $\delta t \ll 1$ employing at each step Lanczos basis (typically $N_L = 10$) 
generating the evolution  $|\Psi_l(t-\delta t) \rangle \to |\Psi_l(t) \rangle$. The described procedure is
very robust and can be easily tested by changing $N_s, \delta E, \delta t$. In the following 
examples we study rather universal (but nontrivial) regime of quite elevated 
$T \sim 5$ where we use $N_s =10,  \delta E=0.01, \delta t=0.01$. Note that for systems considered 
($L=26$) number of basis states is typically $N_{st} \sim 10^7$ so that $\delta E$ is still much bigger
than the average level distance. 

\begin{figure}
\includegraphics[width=0.48\textwidth]{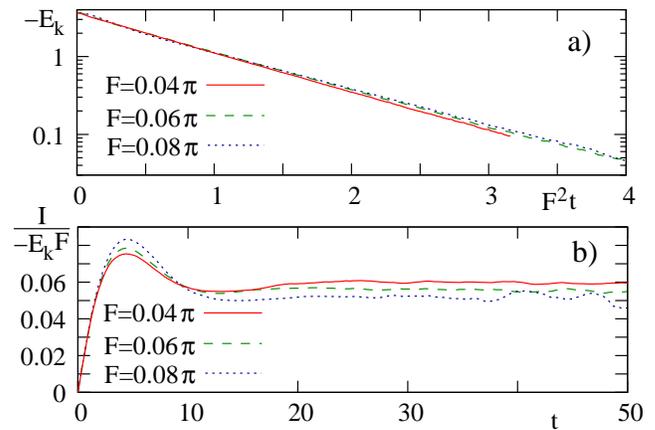}
\caption{(Color online) Real--time response: a) the kinetic energy $E_k(t)$, and b) 
the renormalized current $I(t)/(E_k(t) F)$ obtained for $N=26$, $V=1.4$ and $W=1$. 
The electric field $F$ is switched on at $t=0$ and kept constant for $t>0$.}
\label{fig2}
\end{figure}

\begin{figure}
\includegraphics[width=0.48\textwidth]{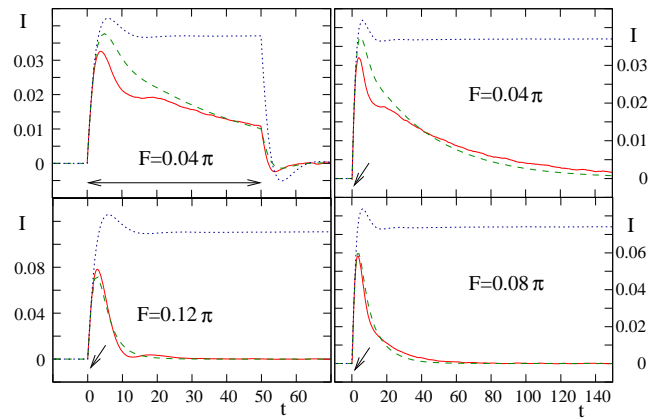}
\caption{(Color online) $I(t)$ for $N=26$, $V=1.4$ and $W=1$.
Arrows indicate the instants of time, when the electric field is switched on (all panels) and off
(upper left panel).  
Full lines represent results obtained via unrestricted numerical solution, whereas dotted 
and dashed lines show approximate results of the LR theory,  $I_{\mathrm{LR}}(t)$, and the extended LR 
approach, $I_{\mathrm{ER}}(t)$, respectively.}
\label{fig3}
\end{figure}

Let us start with the analysis of the non--integrable case where we take $V=1.4$ and $W=1$. 
For such choice of parameters LR $\sigma(\omega)$ \cite{u2}  is rather featureless, 
i.e. broad in $\omega$ and,
therefore, we may expect that numerical results are free of artifacts originating 
from some peculiar features of  $\sigma(\omega)$.
Fig.~\ref{fig2} shows the time dependence of the kinetic energy $E_k(t)$ (panel a) and 
the renormalized current $I(t)/[E_k(t) F]$ (panel b)  for a system driven by a constant field $F(t>0)=F$. 
One can see that $-E_k(t)$ goes  exponentially to zero  
with the decay rate $\propto F^2$ and that the ratio $I(t)/[E_k(t)F]$ is almost constant for longer $t$. 
The dominating nonlinear effect thus consists in the increase of $E_k$. 
Since the latter also represents the sum rule, $\int \sigma(\omega) d\omega \propto |E_k|$, 
we formulate and test the following conjecture: $I(t)$ may
be well approximated by the LR theory extended to account for the increase of $E_k(t)$,
\begin{eqnarray}
I(t) \simeq I_{\mathrm{ER}}(t) &=& \frac{E_k(t)}{E_k(0)} I_{\mathrm{LR}}(t), \label{ELR} \\
I_{\mathrm{LR}}(t)&=&\int_{0}^t \mathrm{d}t' \sigma(t-t') F(t'), \label{ELR1}
\end{eqnarray}
where the conductivity response $\sigma(t-t')$ (evaluated from $\sigma(\omega)$)  
and consequently $I_{\mathrm{LR}}(t)$ are determined within initial equilibrium state.
Eq.~(\ref{ELR}) alone does not allow one to predict the real--time response without solving the
time--dependent  Schr\"odinger equation, since $E_k(t)$ is still needed.
Although not fully evident, the accurate scaling presented in
Fig.~\ref{fig2}a together with Eqs.~(\ref{hdot}) and (\ref{ELR}) 
indicate that the increase of $E_k$ is just proportional 
to the increase of the total energy 
\begin{equation}
\dot{E}_k(t)=\gamma \dot{E}(t)=\gamma L F I_{\mathrm{ER}}(t).
 \label{dote}
 \end{equation} 
Therefore, in the long--time regime of a driving with $F(t) =F$ one obtains  
$I_{\mathrm{ER}}(t)=\sigma_0 F E_k(t)/E_k(0)$ and, 
consequently, $-E_k(t) \propto  \exp(-\alpha F^2 t)$ with $\alpha>0$.    
The coefficient $\gamma$ is independent of $F$ and 
can be estimated from the initial equilibrium state,  
\begin{equation}
\gamma=\frac{E_k(t)-E_k(0)}{E(t)-E(0)}
\simeq \frac{\partial E_k(0)}{\partial T} \left(\frac{\partial E(0)}{\partial T} \right)^{-1}.
\label{ELR2}
\end{equation}
The set of equations  (\ref{ELR}-\ref{ELR2}) fully determines $I_{\mathrm{ER}}(t)$.
Therefore, similarly to the LR theory, the real--time response can be calculated without explicit
solution of the time--dependent problem.  From the numerical point of view,
these equations are not more demanding than the LR theory, as the only extension consists in the
differential equation (\ref{dote}). In Fig.~\ref{fig3} we show the accuracy 
of the approximation given by Eq.~(\ref{ELR}). This figure demonstrates also
 the significance of the increase of total energy or Joule heating as the 
 dominating nonlinear mechanism. Here, the value of $\gamma$ has been determined from full diagonalization of a $10$--site Hamiltonian. 
 
We now turn our attention to the real--time response of an integrable interacting system, 
i.e. $W=0$ case, where the above extension of the LR theory clearly breaks down. 
Now, a relevant reference is the NI $V=0$ system, where
for $F(t)=F$ current and the kinetic energy exhibit Bloch oscillations with $\omega_B= F$.  
It is expected that even at $V>0$ a dissipationless component of current 
will retain the similarity to a NI system. 
We present in the following numerical results for the integrable metallic case
$V=1, W=0$, where $D$ (at $T  \to \infty$) covers approximately half of the total spectral 
weight of $\sigma(\omega)$.  
\begin{figure}
\includegraphics[width=0.48\textwidth]{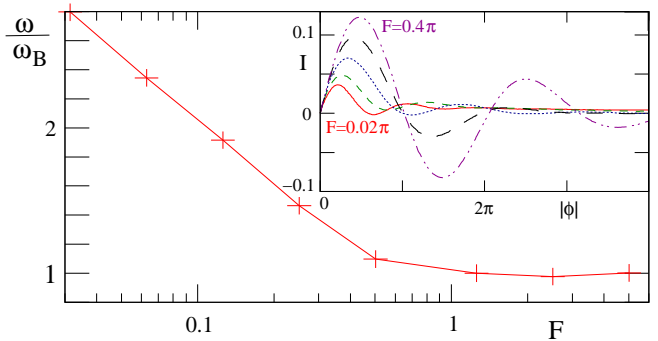}
\caption{(Color online)  Current-oscillation frequency $\omega$ vs. field $F$
normalized to  $\omega_B$ of NI fermions, evaluated  for $N=26$, $V=1, W=0$.
Inset:  $I$ vs. flux $|\phi|$  for  different $F/\pi=0.02, 0.04, 0.08, 0.16,0.4$.}
\label{fig4}
\end{figure}

\begin{figure}
\includegraphics[width=0.3\textwidth]{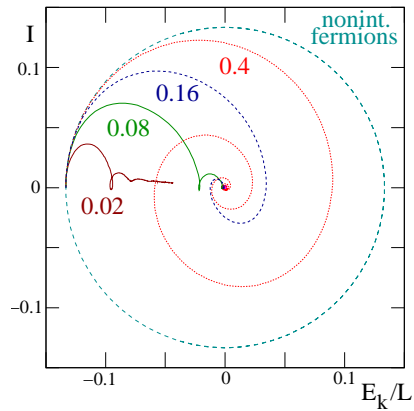}
\caption{(Color online) Relation between current $I(t)$ and kinetic energy density $E_k(t)/L$ 
for parameters  as in Fig.~\ref{fig4}, for different fields $F/\pi= 0.02 - 0.4$. Shown is also the results for NI
fermions.  The point and the arrow mark the initial state and the direction of  evolution, respectively.}
\label{fig5}
\end{figure}

Inset in Fig.~\ref{fig4} shows $I(\phi)$ with $\phi=-F t$
for the integrable $t$-$V$ system driven by a constant electric field $F$. 
Contrary to the previously discussed  $W \neq 0$  case, one observes a strong oscillatory behavior. 
For large enough $F>F^* \sim 0.5$, the frequency of the current oscillations $\omega$
(evaluated here from the second maximum of $I(t)$)
is approximately the same as $\omega_B$ of NI fermions. 
Similar prediction has been obtained within nDMFT (see the discussion of
a metallic phase in  Ref. \cite{DMFT}). 
However, in the regime of a low electric fields $F< F^*$, the correlated system oscillates faster 
than $\omega_B$. Within the investigated $t$-$V$ model,  $\omega/\omega_B$  
is found to increase {\em logarithmically} when the electric field decreases (see Fig.~\ref{fig4}). 
The effect is present also for other $V<2 t_h$ although evidently the logarithmic regime
disappears for $V \to 0$ where the NI universality sets in, i.e. in this case the crossover 
also moves to $ F^* \to 0$.  

At present we are unable to provide a physical 
explanation for the enhancement of $\omega/\omega_B$ or to predict whether 
the  logarithmic  scaling breaks down for very low $F$. At least it is evident that
$F>F^*$ response is similar to NI fermions in that
both $I(t)$ and $E_k(t)=E(t)$ change sign during the evolution (see Fig. \ref{fig5}). On the contrary,
the kinetic energy remains negative in the weak--field regime.

Still, for sufficiently short times of driving $t>0$ the system's state  is almost the initial one, i.e.
$\rho(t)\simeq \rho(0) $, and the changes of $I(t)$ and $E_k(t)$ come 
from their explicit flux dependencies in Eqs.~(\ref{h1}-\ref{jdef}).
Then, 
one can easily find from Eqs. (\ref{hkdot}-\ref{idot})  that $[E_k(t)]^2+[L I(t)]^2 \simeq [E_k(0)]^2$, 
independently of the form of interaction term $H_I$. 
It explains, why  the initial response as well as the strong--field response of a correlated system are 
the same as the response of NI fermions 
\cite{my}. 
Moreover, assuming that the $E_k(t)$ and $I(t)$ do not cross the outer circle in Fig.~\ref{fig5} (what holds true in all investigated cases), one finds 
that the magnitude of current
is bounded from above by the initial kinetic energy. 

 In conclusion, our study reveals several novel features of  isolated  
interacting fermionic systems under
constant or time-dependent driving force. For a generic system of tight--binding 
electrons we present  a simple extension of the LR theory taking into account the change of kinetic energy.
This approach gives a very satisfactory description of numerical results, 
at least for high enough $T$. Here, the
basic condition is that in spite of driving, the system at all times satisfies the quasi-equilibrium
relation between the kinetic energy $E_k(t)$ and the total energy $E(t)$. 
It seems plausible that the necessary condition for such a development is (fast enough) 
relaxation of current, here due to the  Umklapp processes.
It  remains  to be investigated whether a similar behavior applies to interacting systems
beyond the tight-binding description (when e.g. the sum rule for 
$\sigma(\omega)$ is not directly $E_k$)  or in low $T$ regime. In the latter case
the linear dependence $\dot{E}_k(t)\propto \dot{E}(t)$ should probably be replaced with a more general 
one ${E}_k(t)={E}_k[E(t)]$. 

On the other hand, integrable systems show strikingly different development. To first approximation
the current $I(t)$ under constant driving reveals oscillations with the dominant dependence only on the 
flux $\phi(t)$. While at large $F>F^*$ this has clear connection
to the Bloch oscillations of tight-binding electrons with characteristic $\omega \sim \omega_B$,
the effective  $\omega$ is increasing for lower $F<F^*$ while at the same time oscillations are 
becoming damped (in fact the damping is found to be strongest at $F \sim F^*$). One part of
this phenomena, in particular the damping,  can be attributed to the breaking of integrability by 
introducing time--dependent $\phi(t)$. Oscillations themselves in the {\em low--field regime}
are the signature of a coherent behavior which has to be intimately related to finite 
charge stiffness $D(T>0)>0$ since they disappear outside the metallic regime, e.g. for $V>2t_h$ 
at half-filing $n=1/2$.   Alternatively, if the field is switched off in this regime 
before reaching $E_k \sim 0$ then the current relaxes  to a finite $I(t=\infty)>0$ being a direct 
consequence of $D(T>0)>0$. In any case, the observed phenomena reveal that even
in weak fields the response of (near)integrable systems is far from the one expected from the
equilibrium LR theory, but at the same time quite universal  representing an open challenge
for a proper explanation and also possible experimental observation.
  
\acknowledgements

We acknowledge fruitful discussions with J. Bon\v{c}a and X. Zotos.
This work has been support by the Program P1-0044 of the Slovenian Research Agency and
FP6-032980-2 NOVMAG project.

\end{document}